\documentclass[noshowpacs,preprintnumbers,amsmath,amssymb]{revtex4}
\usepackage{amsmath}
\usepackage{amsfonts}
\usepackage{amssymb}
\usepackage[dvips]{graphicx}
\usepackage{color}
\usepackage{graphicx}
\usepackage{bm}
\usepackage{dcolumn}
\usepackage{color}
\usepackage{pst-plot}

\begin{document}

\title{Step-like discontinuities in Bose-Einstein condensates and Hawking radiation: \\ the hydrodynamic
limit
}

\author{Alessandro Fabbri and Carlos Mayoral}
\email{afabbri@ific.uv.es, carlos.mayoral@uv.es}
\affiliation{Departamento de F\'isica Te\'orica and IFIC,\\ Universidad de Valencia-CSIC, C.\ Dr.\ Moliner 50, 46100 Burjassot, Spain}

\date{\today}

\begin{abstract}
We present a detailed analytical analysis of the propagation of Bogoliubov phonons on top of Bose-Einstein condensates with spatial and temporal step-like discontinuities in the speed of sound in the hydrodynamic limit. We focus on some features in the correlations patterns, in particular of density-density correlations. The application to the study of the Hawking signal in the formation of acoustic black hole-like configurations is also discussed.
\noindent
\end{abstract}


\maketitle

\section{Introduction}

The properties of quantum systems in which some parameters experience instantaneous changes in time (or quantum quench, see for instance \cite{kja}-\cite{scc}), have been considered with applications in many areas of physics.

Such step-like discontinuities (in time, and even in space) have been analysed, in particular, for what concerns the propagation of Bogoliubov phonons on top of Bose-Einstein condensates \cite{barcelogaraydopio}-\cite{ftemp} in connection with features of acoustic black holes and cosmology using the so called gravitational analogy \cite{blv}.

Within the same context, our aim is to study correlations (in particular density-density correlations), by developing an approximation scheme to analyse, from an analytic point of view, the Hawking signal in the formation of acoustic black-hole configurations, as considered in \cite{fteo}, \cite{fnum} (see also \cite{cp1}).

In this analysis, restricted to the hydrodynamical approximation, we present a detailed discussion of phonons propagation in backgrounds with step-like discontinuities in the speed of sound, spatial (section 4.1) and temporal (section 4.2) together with some features in the correlations patterns. Then, in section 4.3 we shall combine the two previous cases and repeat the analysis by considering the formation of an inhomogeneous (step-like) configuration out of a homogeneous one. The application to study the Hawking signal from incipient acoustic black holes in the hydrodynamical limit is considered in the final section 5, where it is shown that a structure similar to that found in \cite{fteo} can indeed be recovered .

%
%
\section{\bf The basic equations}

In the dilute gas approximation, a Bose gas can be described \cite{ps} by a field operator $\hat \Psi$ with
equal-time commutator
\begin{equation}\label{etc}
[\hat \Psi (t,\vec x), \hat \Psi^{\dagger}(t,\vec x')]=\delta^3(\vec x- \vec x')
\end{equation}
and time evolution given by
\begin{equation}
i\hbar \partial_t \hat \Psi = \left(-\frac{\hbar^2}{2m}\vec \nabla^2 + V_{ext} + g\hat \Psi^{\dagger}\hat \Psi\right)\hat \Psi\ ,
\end{equation}
where $m$ is the mass of the atoms, $V_{ext}$ the external potential and $g$ the nonlinear atom-atom interaction  constant. By considering linear fluctuations around the classical macroscopic condensate in the form
\begin{equation} \hat \Psi \sim \Psi_0 (1 + \hat \phi)\ ,\end{equation}
 the condensate wavefunction $\Psi_0$ satisfies the Gross-Pitaevski equation
\begin{equation}\label{gp}
i\hbar\partial_t \Psi_0 = \left(-\frac{\hbar^2}{2m}\vec \nabla^2 + V_{ext} + g n \right)\Psi_0\ ,
\end{equation}
with $n=|\Psi_0|^2$ the number density, while the Bogoliubov-de Gennes equation for the fluctuations  takes the form
\begin{equation}\label{bdg}
i\hbar  \partial_t   \hat \phi=  \left( -\frac{\hbar^2}{2m}\vec \nabla^2 - \frac{\hbar^2}{m}\frac{\vec \nabla \Psi_0 }{\Psi_0} \vec \nabla\right)\hat\phi +mc^2 (\hat\phi + \hat\phi^{\dagger})\ ,
\end{equation}
the speed of sound being  $c=\sqrt{\frac{gn}{m}}$.

\section{Our model}

We shall consider condensates of constant density $n$ and velocity $v$ (for simplicity along one dimension, say $x$). As explained in \cite{fnum}, nontrivial configurations are still possible provided one varies
the coupling constant $g$ (and therefore $c$) and the external potential but keep the sum $gn+ V_{ext}$ constant. In this way the
plane-wave $\Psi_0=\sqrt{n}e^{ik_0x-iw_0t}$, where $v=\frac{\hbar k_0}{m}$ is the condensate velocity and $\hbar w_0=\frac{mv^2}{2}+ gn+ V_{ext}$,  is a solution of (\ref{gp}) everywhere. The transverse degrees
of freedom are assumed to be frozen.


%

To consider the hydrodynamic limit it is useful to introduce the two hermitean operators
$\hat n^1$ (density fluctuation) and $\hat \theta^1$ (phase fluctuation) via
\begin{equation}\label{dph}
\hat\phi = \frac{\hat n^1}{2n} + i\frac{\hat\theta^1}{\hbar}\ . \end{equation}
Inserting in (\ref{bdg}) and expanding as
\begin{equation}
\label{dpm}
\begin{array}{c}
\hat n^1 (t,x) =\sum_j \left[ \hat a_j n^1_j (t,x) + \hat a_j^{\dagger} n^1_j(t,x)^* \right]\ ,\\
\hat \theta^1 (t,x) =\sum_j \left[ \hat a_j \theta^1_j (t,x) + \hat a_j^{\dagger} \theta^1_j(t,x)^* \right]\ ,
\end{array}
\end{equation}
we get the mode equations
\begin{equation}
\label{dpf}
\begin{array}{c}
(\partial_t +v\partial_x)n^1 +\frac{n}{m}\partial_x^2\theta^1=0\ ,\\
(\partial_t +v\partial_x)\theta^1 + \frac{mc^2}{n}n^1 -\frac{\hbar^2}{4mn}\partial_x^2 n^1 =0\ .
\end{array}
\end{equation}
The equal time commutator (\ref{etc}) becomes, in this representation,
 \begin{equation}\label{etct} [\hat n^1 (t,x), \hat\theta^1(t,x')]=i\hbar\delta(x-x')\ ,\end{equation}
which after integration gives the modes normalization
\begin{equation}
\label{nord}
\int dx [n^1_j \theta_{j'}^{1*} - n_j^{1*} \theta^1_{j'}]=i \delta_{jj'}\ .\end{equation}

 \section{Hydrodynamics}

In the hydrodynamical approximation one probes distances much larger than the healing length $\xi=\frac{\hbar}{mc}$. This limit is easily achieved in the density-phase representation (\ref{dph}),
where in the second of eqs. (\ref{dpf}) we consider \begin{equation} \frac{mc^2}{n}(n^1 -\frac{\xi^2}{4}\partial_x^2 n^1)
\sim \frac{mc^2}{n}n^1\ , \end{equation}
leading to
\begin{equation}
\label{hyd}
\begin{array}{c}
(\partial_t +v\partial_x)n^1 +\frac{n}{m}\partial_x^2\theta^1=0\ ,\\
(\partial_t +v\partial_x)\theta^1 + \frac{mc^2}{n}n^1 =0\ .
\end{array}
\end{equation}
Eqs. (\ref{hyd})
can be rewritten as a single second order equation for $\theta^1$ provided one algebraically extracts $n^1$
from the second
\begin{equation}\label{nm}
n^1 = - \frac{n}{mc^2}(\partial_t +v\partial_x)\theta^1
\end{equation}
and substitutes it in the first to get the second order differential equation
\begin{equation}\label{sde}
-(\partial_t+v\partial_x)\frac{1}{c^2}(\partial_t+v\partial_x)\theta^1 + \partial_x^2\theta^1=0\ .
\end{equation}
In the language of the gravitational analogy \cite{blv}, $\theta^1$ becomes a massless and minimally coupled scalar field propagating in the curved acoustic geometry \begin{equation}
\label{acm}
g_{\mu\nu}=\frac{n}{mc}\left(
  \begin{array}{cccc}
   -(c^2-v^2)  & -v & 0 & 0 \\
     -v & 1 & 0 & 0  \\
     0  & 0 & 1 & 0  \\
     0  & 0 & 0 & 1 \\
  \end{array}
\right)
\end{equation}
and satisfying the Klein-Gordon equation
\begin{equation}\label{kgq}
\Box\theta^1 = \frac{1}{\sqrt{-g}}\partial_\mu\left[\sqrt{-g}g^{\mu\nu}\partial_\nu \theta^1\right]=0\ ,
\end{equation}
which corresponds (we remind that in our model $n$ and $v$ are constant) to eq. (\ref{sde})
provided $\theta^1$ does not depend upon the transverse dimensions.
The scalar product is given by, see eqs (\ref{nord}) and (\ref{nm}),
\begin{equation}
\label{scp}
(\theta^1_j,\theta^1_{j'})=i\int dx \frac{n}{mc^2}\left[\theta^{1*}_j(\partial_t+v\partial_x)\theta^1_{j'}
-\theta^1_{j'}(\partial_t+v\partial_x)\theta_j^{1*}\right]=\delta_{jj'}\ .
\end{equation}
We shall consider the simplest possible nontrivial configurations, namely step-like discontinuities
of the speed of sound $c$ in both $x$ and $t$. In order to explicitly solve these problems we shall
rewrite eq. (\ref{sde}) in the more appropriate form
\begin{equation}\label{seceq}
-\partial_t \left[ \frac{\partial_t \theta^1}{c^2}+ \frac{v}{c^2}\partial_x\theta^1\right] +\partial_x
\left[(1-\frac{v^2}{c^2})\partial_x\theta^1
-\frac{v}{c^2}\partial_t\theta^1\right]=0\ .\end{equation}
Such configurations allow us to extract analytical results which can be
meaningfully compared with the numerical analysis of more complicated regular configurations.\\
To clarify the presence of the $\frac{1}{c^2}$ inside the square brackets in (\ref{seceq}), from (\ref{kgq}) we note that our four dimensional acoustic metric (\ref{acm}) has $\sqrt{-g}\sim \frac{1}{c}$ and that the inverse metric is
\begin{equation}
g^{\mu\nu}=\frac{m}{nc}\left(
  \begin{array}{cccc}
   -1 & -v & 0 & 0 \\
     -v & c^2-v^2 & 0 & 0  \\
     0 & 0 & c^2 & 0  \\
     0 & 0 & 0 & c^2 \\
  \end{array}\right),
\end{equation}
i.e., the relevant $t,r$ components carry an additional factor of $\frac{1}{c}$.

\subsection{\bf Step-like discontinuity in $x$ (stationary case)}
\bigskip

Let us consider a surface (say, $x=0$) separating two semi-infinite homogeneous condensates
with different sound speeds, i.e. we consider $c(x)=c_l\theta(-x) + c_r \theta(x)$.
We shall consider $v<0$, that is the fluid flows from right to left.

To solve eq. (\ref{seceq}) in all space we need first to solve it separately in the left ($x<0$) and right ($x>0$) regions, then impose the appropriate matching conditions  at the discontinuity.
Looking at (\ref{seceq}) these are \begin{equation}\label{mcondsec}
\left[\theta^1\right]=0, \ \ \ \
\left[(1-\frac{v^2}{c^2})\partial_x\theta^1
-\frac{v}{c^2}\partial_t\theta^1\right]=0\ ,
\end{equation}
where $[\ ]$ means variation across
the jump at $x=0$.

Eq. (\ref{sde}) is easily solved in left and right regions in terms of plane-waves
$B e^{-i\omega t+ikx}$ with the standard linear dispersion relation $(w-vk)^2=(ck)^2$.
Being the model stationary the matching conditions
(\ref{mcondsec})
are solved at fixed $w$. This motivates the use of the $w$ expansion for the operator
$\hat \theta^1$
\begin{equation}\label{decompw}
\hat \theta^1 (t,x) =\int_0^{+\infty} dw \left[ \hat a_w^u u_w^u (t,x) +  \hat a_w^v u_w^v(t,x) +h.c. \right]\,
\end{equation}
in terms of right-moving \begin{equation}u_w^u=B_u e^{-iwt+ik_ux},\  k_u=\frac{w}{v+c}\ ,\end{equation} and left-moving  waves \footnote{The $u,v$ notation is a relativistic one, where $u_w^u \equiv B_u e^{-iwu}$ and $u_w^v\equiv B_v e^{-iwv}$ are written using the null retarded and advanced coordinates $u=t-\frac{x}{v+c}, v=t-\frac{x}{v-c}$.}
\begin{equation} u_w^v=B_v e^{-iwt+ik_vx},\ k_v=\frac{w}{v-c}\ . \end{equation}

These modes are
ortho-normalized according to
\begin{equation}
\label{eq:norma}
(u_w,u_{w'})=i\int dx \frac{n}{mc^2}\left[u^*_w(\partial_t+v\partial_x)u_{w'}-u_{w'}(\partial_t+v\partial_x)
u^{*}_w\right]=
\delta(w -w')\ ,
\end{equation}

and for infinite homogeneous condensates we have
\begin{equation} |B_u|^2=|B_v|^2=\frac{mc^2}{4\pi n}\left|\frac{\frac{dk}{dw}}{\Omega }\right|=\frac{mc}{4\pi nw}\ ,\end{equation}
where $\Omega=w-kv$ is the frequency measured in the atoms frame. \\
Our derivation of (\ref{eq:norma}) is based on the spatial integration of the commutator (\ref{etct}), which is time independent. To further check that this scalar product does not depend on time we consider
\begin{equation}\label{tind}
(u_{\omega},u_{\omega'})_{t_2}-(u_{\omega},u_{\omega'})_{t_1}=i\frac{n}{m}\int_{t_1}^{t_2}dt\int dx\partial_t \left[u^*_w\frac{1}{c^2}(\partial_t+v\partial_x)u_{w'}-u_{w'}\frac{1}{c^2}(\partial_t+v\partial_x)
u^{*}_w\right].
\end{equation}
Using the equation of motion (\ref{seceq}), the above expression can be rewritten as
\begin{equation}
\frac{n}{m}\int_{t_1}^{t_2}dt\int dx \left[\frac{v}{c^2}\left(\partial_t u_{\omega}^*\partial_xu_{\omega^{'}}-\partial_xu_{\omega}^*\partial_t u_{\omega^{'}}\right)+u^*_{\omega}\partial_x\left((1-\frac{v^2}{c^2})\partial_xu_{\omega^{'}}-\frac{v}{c^2}\partial_t u_{\omega^{'}}\right)-
u_{\omega^{'}}\partial_x\left((1-\frac{v^2}{c^2})\partial_xu^*_{\omega}-\frac{v}{c^2}\partial_t u^*_{\omega}\right)\right].
\end{equation}
Integration by parts in the third and fourth terms
and use of the boundary conditions (\ref{mcondsec}) for the surface terms shows that this quantity indeed vanishes.\\
In the l region ($x<0$) the generic solution to (\ref{sde}) at fixed $w$ is thus written in the form
\begin{equation}
\label{eq:fieldl}
 \theta^{1}_l=e^{-iwt}\sqrt{\frac{mc_l}{4\pi wn}}
 \left[ e^{i\frac{\omega }{v-c_l}x} A_v^l+
e^{i\frac{\omega }{v+c_l}x}  A_u^l \right] \ ,  \end{equation}
 whereas in the r ($x>0$) region
\begin{equation}\label{eq:fieldr}
 \theta^{1}_r=
e^{-iwt}\sqrt{\frac{mc_r}{4\pi wn}}
\left[ e^{i\frac{\omega }{v-c_r}x}  A_v^r+
e^{i\frac{\omega }{v+c_r}x}  A_u^r \right] \ .
\end{equation}
We shall now impose the matching conditions (\ref{mcondsec}) at $x=0$.  They read
\begin{equation}
\label{eq:momentos}
\begin{array}{c}
\sqrt{c_l} ( A_v^l+  A_u^l) = \sqrt{c_r} ( A_v^r+  A_u^r )\ ,
\\
\sqrt{c_l}\{ \left[(1-\frac{v^2}{c_l^2})\frac{1}{v-c_l}+\frac{v}{c_l^2}\right]A_v^l
+ \left[(1-\frac{v^2}{c_l^2})\frac{1}{v+c_l}+\frac{v}{c_l^2}\right]A_u^l\} =\\
=\sqrt{c_r}\{\left[(1-\frac{v^2}{c_r^2})\frac{1}{v-c_r}+\frac{v}{c_r^2}\right]A_v^r
+ \left[(1-\frac{v^2}{c_r^2})\frac{1}{v+c_r}+\frac{v}{c_r^2}\right]A_u^l\} \ .
\end{array}
\end{equation}
They can be rewritten in matrix form
\begin{equation}
\label{eq:matching}
W_l\left(
     \begin{array}{c}
       A_v^l \\
       A_u^l \\
     \end{array}
   \right)=W_r\left(
                \begin{array}{c}
                  A_v^r \\
                  A_u^r \\
                \end{array}
              \right)
\end{equation}

or more explicitly

\begin{equation}
\label{eq:matchingexplicito}
\sqrt{c_l}\left(
  \begin{array}{cc}
   1  & 1 \\
     -\frac{1}{c_l} & \frac{1}{c_l} \\
  \end{array}
\right)
\left(
     \begin{array}{c}
       A_v^l \\
       A_u^l \\
     \end{array}
   \right)=\sqrt{c_r}\left(
  \begin{array}{cc}
   1  & 1 \\
   -\frac{1}{c_r}  & \frac{1}{c_r} \\
  \end{array}
\right)\left(
                \begin{array}{c}
                  A_v^r \\
                  A_u^r \\
                \end{array}
              \right) \ .
\end{equation}

By inverting $W_l$ we get  \begin{equation}
\label{eq:matching}
     \left( \begin{array}{c}
       A_v^l \\
       A_u^l \\
     \end{array} \right)
   =M_{scat} \left(
                \begin{array}{c}
                  A_v^r \\
                  A_u^r \\
                \end{array}
              \right) \ ,
\end{equation}
the scattering matrix $M_{scat}\equiv W_l^{-1} W_r$ being
 \begin{equation}
\label{eq:scattering}
M_{scat}=\frac{1}{2\sqrt{c_rc_l}}\left(
            \begin{array}{cc}
               c_r+c_l &
               c_r-c_l \\
               c_r-c_l &
               c_r+c_l \\
            \end{array}
          \right) \ .
\end{equation}

\subsubsection{\bf `in' and `out' basis and two-point function}

\begin{figure}[htbp]
\begin{center}

\resizebox{!}{9cm}{\includegraphics{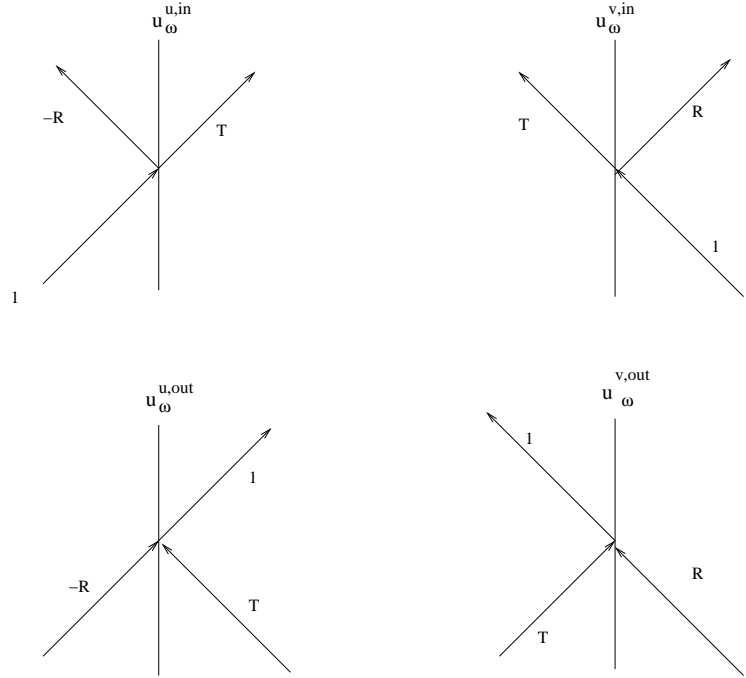}}

\caption{`in' and `out' modes.}
\label{figure:modosin}
\end{center}
\end{figure}

Let us construct the `in' modes ,  see Figure \ref{figure:modosin}.

\noindent These are initially left moving modes from the right ($u_{\omega}^{v,in}$)
and initially  right-moving ones from the left ($u_{\omega}^{ u,in}$).

{\center{\bf Modes \textbf{$u_{\omega}^{ v,in}$}}}

Consider a left-moving ($v$) wave from the right (r) region with unit amplitude, i.e.
$A_v^r=1$. The scattering at $x=0$ will produce a reflected wave (right-moving $u$ wave in the $r$
region, with amplitude $A_u^r$) plus a transmitted one (left-moving in the left(l) region, $A_v^l$).
Thus we have to solve the following system of equations
\begin{equation}
\label{scattamp}
     \left( \begin{array}{c}
       A_v^l \\
       0  \\
     \end{array} \right)
   = \frac{1}{2\sqrt{c_r c_l}}\left(
            \begin{array}{cc}
               c_r+c_l &
               c_r-c_l \\
               c_r-c_l &
               c_r+c_l \\ \end{array} \right) \left(
                \begin{array}{c}
                  1  \\
                  A_u^r \\
                \end{array}
              \right) \ .
\end{equation}
Its solutions are \begin{equation} \label{erreti}A_u^r= - \frac{c_r-c_l}{c_r+c_l}\equiv R\ ,\\ A_v^l
= \frac{2\sqrt{c_rc_l}}{c_r+c_l}\equiv T\ ,\end{equation}

satisfying the unitarity condition
\begin{equation}
|A_u^r|^2 + |A_v^l|^2\equiv |R|^2
+ |T|^2=1\ .\end{equation}

We note that this construction is valid only when the flow is everywhere subsonic.
Indeed, if the left region were supersonic, $|v|>c_l$, we would have two transmitted components instead of one because there also the `right-moving' one ($A_u^l$) would be dragged to the left by the fluid (i.e. $c_l+v<0$). Mathematically the system does not admit a unique solution since we would have three amplitudes to determine with only two equations. From the physical point of view we would need a smoother configuration, an acoustic black hole,  with a transition
region from the right to the left regions  of nonvanishing width $\Delta \sigma_x$ (see the final section 5).
{\center{\bf Modes \textbf{$u_{\omega}^{ u,in}$}}}

A right-moving ($u$) wave from the l region with unit amplitude, i.e.
$A_u^l=1$, gives rise to a reflected wave (left-moving $v$ wave in the $l$
region, with amplitude $A_v^l$) plus a transmitted one (right-moving in the r region, $A_u^r$).

The amplitudes are obtained by solving
\begin{equation}
\label{modeuin}
     \left( \begin{array}{c}
       A_v^l \\
       1  \\
     \end{array} \right)
   = \frac{1}{2\sqrt{c_r c_l}}\left(
            \begin{array}{cc}
               c_r+c_l &
               c_r-c_l \\
               c_r-c_l &
               c_r+c_l \\ \end{array} \right) \left(
                \begin{array}{c}
                  0  \\
                  A_u^r \\
                \end{array}
              \right) \ ,
\end{equation}\label{erremenoti}
that gives \begin{equation} A_v^l=\frac{c_r-c_l}{c_r+c_l}= -R\ ,  A_u^r=\frac{2\sqrt{c_rc_l}}{c_r+c_l}=T\ .\end{equation}
Again, this construction is for subsonic flows. Were the left $l$ region supersonic, the initial $u$ mode could not propagate to the right, but would be dragged to the left by the fluid.\\

Similarly, one can construct the `out' modes $u_{\omega}^{v,out}, u_{\omega}^{u,out}$, which are represented in Figure \ref{figure:modosin}.
\\

The relations between the `in' and `out' modes
\begin{eqnarray}
\label{eq:relationinoutmo}
  u_{\omega}^{v,in} &=& T u_{\omega}^{ v,out}+R u_{\omega}^{ u,out}\ , \nonumber\\
  u_{\omega}^{ u,in} &=& -R u_{\omega}^{v,out}+T u_{\omega}^{u,out}\ ,
\end{eqnarray}
allow us to determine the relation between the $\hat a, \hat a^{\dagger}$ operators in the `in' and 'out' decompositions
\begin{equation}
\hat \theta^1 (t,x) =\int_0^{+\infty} dw \left[ \hat a_w^{v,in(out)} u^{v,in(out)}_w (t,x)+ \hat a_w^{u,in(out)} u^{ u,in(out)}_w (t,x)  + h.c. \right]\ ,
\end{equation}
i.e.
\begin{eqnarray}\label{eq:relationinoutop} \hat a_{\omega}^{v,out} &=& T \hat a_{\omega}^{v,in}-R \hat a_{\omega}^{u,in}\ , \nonumber \\ \hat a_{\omega}^{u,out} &=& R\hat a_{\omega}^{v,in}+T\hat a_{\omega}^{u,in} \ . \end{eqnarray}
Since (\ref{eq:relationinoutop}) only involve annihilation operators the vacuum state is unique $ |in\rangle=|out\rangle $.


We can then write down the symmetric two-point function
\begin{eqnarray} \label{tupi}
\langle in| \{ \hat\theta^1 (t,x),\hat\theta^1 (t',x')\} |in\rangle \equiv
\frac{1}{2}\ \langle in| \hat\theta^1 (t,x)\hat\theta^1 (t',x')
+ \hat\theta^1 (t',x')\hat\theta^1 (t,x) |in\rangle=
\ \ \ \ \ \ \ \ \ \ \ \ \ \  \ \ \ \ \ \  \nonumber \\ \frac{\hbar}{2} \int_0^{+\infty} dw\Big[ u^{v,in(out)}_w (t,x)u^{v,in(out)*}_w(t',x') +  u^{u,in(out)}_w (t,x)u^{u,in(out)*}_w(t',x') +c.c.\Big]\ ,
  \ \ \ \ \ \ \ \ \ \ \ \ \ \ \ \ \ \ \   \nonumber  \end{eqnarray}
where,
explicitly,
\begin{eqnarray}\label{uvin}
&u^{v,in}_w=u_{\omega}^{v,r}(t,x)+Ru_{\omega}^{u,r}(t,x)+ Tu_{\omega}^{v,l}(t,x)\ ,&\nonumber\\
&u^{u,in}_w=u_{\omega}^{u,l}(t,x)-Ru_{\omega}^{v,l}(t,x)
+Tu_{\omega}^{u,r}(t,x)\ ,&\end{eqnarray}
with left and right moving components in the $l$ and $r$ regions
\begin{eqnarray}
\label{modes}
  u_w^{u,l}(t,x) &=& \sqrt{\frac{mc_l}{4\pi\omega n}}e^{-i\omega t+i\frac{\omega}{v+c_l}x}, \,\,\,\,u_w^{u,r}(t,x)=u_w^{u,l}(t,x)[c_l\rightarrow c_r]\nonumber\\
  u_w^{v,l}(t,x) &=& \sqrt{\frac{mc_l}{4\pi\omega n}}e^{-i\omega t+i\frac{\omega}{v-c_l}x}, \,\,\,\,u_w^{v,r}(t,x)=u_v^l(t,x)[c_l\rightarrow c_r]\ .
\end{eqnarray}

Considering for instance points situated, respectively, in the l ($x<0$) and r ($x'>0$) regions we have
\begin{eqnarray}
\label{eq:twopoint1}
\langle in| \{ \hat\theta^1 (t,x),\hat\theta^1 (t',x')\} |in\rangle=
\frac{\hbar}{2} \int_0^{\infty}d\omega \Big[ T^*u_w^{u,l}(t,x)u_w^{u,r *}(t',x') + \nonumber \\ T  u_w^{v,l}(t,x)u_w^{v,r *}(t',x') +  (R^*T-RT^*)u_{\omega}^{v,l}(t,x)u_{\omega}^{u,r *}(t',x')  + c.c.\Big] = \nonumber \\
-\frac{\hbar mc_rc_l}{2\pi n(c_r+c_l)} \ln [(t-t'-\frac{x}{v+c_l}+\frac{x'}{v+c_r})
(t-t' -\frac{x}{v-c_l}+\frac{x'}{v-c_r})]  \ .
\end{eqnarray}
The non vanishing contributions are of $uu$ and $vv$ type, while the $uv$ correlations, being proportional to $(R^*T-RT^*)$, vanish.\\
In section III.A of \cite{nostrodisp} we have considered dispersion effects for spatial step-like discontinuities in subsonic flows. There it is shown that the $\xi \rightarrow 0$ limit is smooth and that the hydrodynamic modes give the main contribution to the correlators.\\
Finally, we stress that from the point of view of the gravitational analogy the nontrivial scattering structure here outlined depends crucially on the fact that the geometry (\ref{acm}) is  four-dimensional. Were $g_{\mu\nu}$ two-dimensional, the field $\theta^1$ would be
conformal and there would be no scattering between right-moving and left-moving components.

\bigskip
{\subsection{\bf Step-like discontinuity in $t$ (homogeneous case)}}
\bigskip

The other case of interest is the step-like discontinuity in $t$ (say, at $t=0$), separating
two infinite homogeneous condensates: $c(t)=c_{in}\theta(-t)+c_{out}\theta(t)$.

To solve (\ref{seceq}) for all times means we need to write down the general solutions in both
`in' ($t<0$) and `out' ($t>0$) regions and then impose the appropriate matching conditions for the
problem at $t=0$, which are (see (\ref{seceq}))
\begin{equation}\label{mcondsecb}
\left[\theta^1\right]=0, \ \ \ \
\left[ \frac{\partial_t \theta^1}{c^2}+ \frac{v}{c^2}\partial_x\theta^1\right]=0\ .
\end{equation}

In both `in' and `out' regions the solutions are plane-waves $e^{-iwt+ikx}$, and being the problem
homogeneous the boundary conditions (\ref{mcondsecb}) are solved at fixed $k$. Therefore we shall use the
$k$ decomposition for $\hat \theta^1$
\begin{equation}\label{decompk}
\hat \theta^1 (t,x) =\int_{-\infty}^{+\infty} dk \left[ \hat a_k u_k (t,x)  + \hat a_k^{\dagger} u_k(t,x)^* \right]\ ,
\end{equation}
where the positive-frequency modes \begin{equation} \label{mow} u_k=C e^{-iw(k)t+ikx}\  \end{equation}   are right-moving, $w_u=(v+c)k$, when $k>0$, and left-moving, $w_v=(v-c)k$, as $k<0$. Such modes are
ortho-normalized according to
\begin{equation}
\label{eq:norm}
(u_k,u_{k'})=i\int dx \frac{n}{mc^2}\left[u^*_k(\partial_t+v\partial_x)u_{k'}-u_{k'}(\partial_t+v\partial_x)
u^{*}_k\right]=
\delta(k -k')\ ,
\end{equation}
i.e.
\begin{equation}
|C|^2=\frac{mc^2}{4\pi |\Omega | n}=\frac{mc}{4\pi |k| n}\ .
\end{equation}
This scalar product too is time independent,
i.e.
\begin{equation}\label{tind}
(u_{k},u_{k'})_{t_2}-(u_{k},u_{k'})_{t_1}=i\frac{n}{m}\int_{t_1}^{t_2}dt\int dx\partial_t \left[u^*_k\frac{1}{c^2}(\partial_t+v\partial_x)u_{k'}-u_{k'}\frac{1}{c^2}(\partial_t+v\partial_x)
u^{*}_k\right] =0\ .\end{equation}
For $t_1t_2>0$ this is trivial, whereas for $t_1t_2<0$ this is guaranteed  by the boundary conditions (\ref{mcondsecb}) at the timelike discontinuity. See also \cite{wswav}.

Note that the general solution at fixed $k$
\begin{equation}
\theta^1=C e^{ikx} \left[ e^{-ik(v-c)t} A_v + e^{-ik(v+c)t} A_u \right]\ \end{equation}
 contains a positive-frequency and a negative-frequency component: when $k>0$ ($k<0$) it describes a positive frequency right-moving (left-moving) mode and a negative frequency left-moving (right-moving) one.

In the `in' region we write
\begin{equation}
\theta^1_{in}= e^{ikx}\sqrt{\frac{mc_{in}}{4\pi |k| n}} \left[ A_v^{in}e^{-ik(v-c_{in})t}  + A_u^{in}e^{-ik(v+c_{in})t}  \right],
\end{equation}
whereas in the `out' region
\begin{equation}
\theta^1_{out}= e^{ikx}\sqrt{\frac{mc_{out}}{4\pi |k| n}} \left[ A_v^{out}e^{-ik(v-c_{out})t}  + A_u^{out}e^{-ik(v+c_{out})t}  \right].
\end{equation}
The matching conditions (\ref{mcondsecb}) read

\begin{equation}
\label{eq:momentost}
\begin{array}{c}
\frac{1}{c_{out}^{3/2}}\{ [ (v-c_{out}) - v] A_v^{out}
+ [ (v+c_{out}) - v] A_u^{out}\} =\\ \frac{1}{c_{in}^{3/2}}\{ [ (v-c_{in}) - v] A_v^{in}
+ [ (v+c_{in}) - v] A_u^{in}\}
 \ .
\end{array}
\end{equation}
In matrix form
\begin{equation}
\label{eq:bogo}
W_{out}\left(
     \begin{array}{c}
       A_v^{out} \\
       A_u^{out} \\
     \end{array}
   \right)=W_{in}\left(
                \begin{array}{c}
                  A_v^{in} \\
                  A_u^{in} \\
                \end{array}
              \right)
\end{equation}

or more explicitly

\begin{equation}
\label{eq:matchingexplicitot}
\frac{1}{\sqrt{c_{out}}}\left(
  \begin{array}{cc}
   c_{out}  & c_{out} \\
     -1 & 1 \\
  \end{array}
\right)
\left(
     \begin{array}{c}
       A_v^{out} \\
       A_u^{out} \\
     \end{array}
   \right)=\frac{1}{\sqrt{c_{in}}}\left(
  \begin{array}{cc}
   c_{in}  & c_{in} \\
   -1  & 1 \\
  \end{array}
\right)\left(
                \begin{array}{c}
                  A_v^{in} \\
                  A_u^{out} \\
                \end{array}
              \right) \ .
\end{equation}

By inverting $W_{out}$ we get  \begin{equation}
\label{eq:matchingt}
     \left( \begin{array}{c}
       A_v^{out} \\
       A_u^{out} \\
     \end{array} \right)
   =M_{bog} \left(
                \begin{array}{c}
                  A_v^{in} \\
                  A_u^{in} \\
                \end{array}
              \right) \ ,
\end{equation}
the Bogoliubov matrix $M_{Bog}\equiv W_{out}^{-1} W_{in}$ being
 \begin{equation}
\label{eq:bogoliubov}
M_{Bog}=\frac{1}{2\sqrt{c_{in}c_{out}}}\left(
            \begin{array}{cc}
               c_{in}+c_{out} &
               c_{in}-c_{out} \\
               c_{in}-c_{out} &
               c_{in}+c_{out} \\
            \end{array}
          \right) \ .
\end{equation}

\bigskip
{\subsubsection{\bf Connecting the `in' and `out' basis}}
\bigskip

\begin{figure}[htbp]
\begin{center}

\resizebox{!}{8cm}{\includegraphics{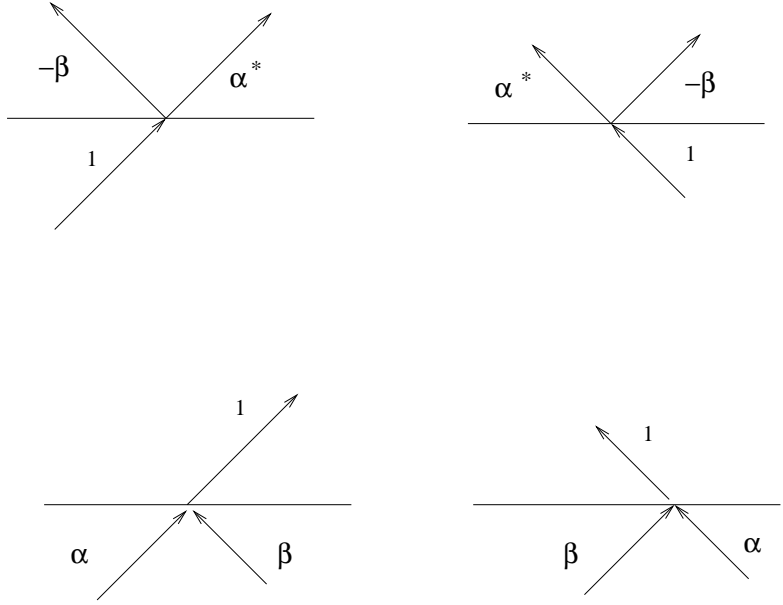}}

\caption{`in' and `out' modes.}
\label{figure:tempgrad}
\end{center}
\end{figure}

The `in' and `out' basis are easily identified.
They are constructed in terms of `in' and `out' positive-frequency modes which are left-moving $u_k^{in(out)}$,\ $k<0$, and right-moving $u_k^{in(out)}$,\ $k>0$.
To connect them, as depicted in Fig.\ref{figure:tempgrad} for subsonic flows, we use the Bogoliubov matrix (\ref{eq:bogoliubov}). Were the `out' region supersonic, both the modes would move to the left. \\

Let us construct explicitly the `in' modes.

{\center{\bf Modes \textbf{$u_{k}^{in}\ (k<0) $}}}
\bigskip\\
A left moving `in' mode  has $A_v^{in}=1,\ A_u^{in}=0$. The coefficients $A_v^{out}$ and $A_u^{out}$ are found by solving
\begin{equation}
\label{bogv}
     \left( \begin{array}{c}
       A_v^{out} \\
       A_u^{out}  \\
     \end{array} \right)
   = \frac{1}{2\sqrt{c_{in} c_{out}}}\left(
            \begin{array}{cc}
               c_{in}+c_{out} &
               c_{in}-c_{out} \\
               c_{in}-c_{out} &
               c_{in}+c_{out} \\ \end{array} \right) \left(
                \begin{array}{c}
                  1  \\
                  0 \\
                \end{array}
              \right) \ .
\end{equation}
The solution is \begin{equation} A_v^{out}=\frac{c_{in}+c_{out}}{2\sqrt{c_{in}c_{out}}}\equiv \alpha^*,\  A_u^{out}=\frac{c_{in}-c_{out}}{2\sqrt{c_{in}c_{out}}}\equiv -\beta\ .\end{equation}
 These coefficients satisfy the unitarity condition
\begin{equation} |A_v^{out}|^2 - |A_u^{out}|^2\equiv |\alpha|^2 - |\beta|^2 = 1\ ,\end{equation}
the $-$ sign meaning that the $A_u^{out}$ are associated to negative frequency and negative norm modes.

\bigskip
{\center{\bf Modes \textbf{$u_{k}^{in}\ (k>0) $}}}
\bigskip\\

A right-moving `in' mode has
$A_v^{in}=0,\ A_u^{in}=1$. From

\begin{equation}
\label{bogu}
     \left( \begin{array}{c}
       A_v^{out} \\
       A_u^{out}  \\
     \end{array} \right)
   = \frac{1}{2\sqrt{c_{in} c_{out}}}\left(
            \begin{array}{cc}
               c_{in}+c_{out} &
               c_{in}-c_{out} \\
               c_{in}-c_{out} &
               c_{in}+c_{out} \\ \end{array} \right) \left(
                \begin{array}{c}
                  0  \\
                  1 \\
                \end{array}
              \right) \ .
\end{equation}
  we get \begin{equation} \label{alfabeta} A_v^{out}=\frac{c_{in}-c_{out}}{2\sqrt{c_{in}c_{out}}}= -\beta,\ \  A_u^{out}=\frac{c_{in}+c_{out}}{2\sqrt{c_{in}c_{out}}}= \alpha^*\ .\end{equation}

The `out' modes can be similarly constructed, see Fig \ref{figure:tempgrad}.
\bigskip

The relation between the `in' and `out' modes can be written compactly
\begin{equation}
\label{iuppi}
  u_{k}^{in} = \alpha^* u_{k}^{out}-\beta u_{-k}^{out *}\ , \nonumber\\
\end{equation}
and it allows us to determine the relation between the annihilation and creation operators in the `in' and `out' decompositions
\begin{equation}
\hat \theta^1 (t,x) =\int_{-\infty}^{+\infty} dk \left[ \hat a_k^{in(out)} u^{in(out)}_k (t,x)+ h.c. \right]\ ,
\end{equation}
i.e.
\begin{equation}\label{eq:relationinoutopk} \hat a_{k}^{out} = \alpha^* \hat a_{k}^{in}-\beta^* \hat a_{-k}^{in\dagger}\ .\end{equation}
 We see that (\ref{eq:relationinoutopk}) involves both annihilation and creation operators, so the two decompositions are inequivalent and in particular their vacuum states are different $|in\rangle\neq |out\rangle $. Up to a normalization factor these two states are related by the formula
 \begin{equation}\label{vacio}
 |in\rangle \sim e^{-\frac{\beta^*}{2\alpha}\int dk' \hat a^{\dagger}_{k'}\hat a^{\dagger}_{-k'}}|out\rangle \ ,
\end{equation}
showing that the $|in\rangle$ state contains correlated pairs of `out' particles with opposite momenta (i.e. one
right-moving and the other left-moving). These features show up in the correlations.

Again, we point out that in the language of the gravitational analogy (\ref{kgq}) all the nontrivial features here discussed are a consequence of the four-dimensional character of the acoustic metric, and that they would be absent if $g_{\mu\nu}$ were two-dimensional and, consequently, the field $\theta_1$ conformal.

\bigskip
{\subsubsection{\bf Two-point function and density-density correlations}}
\bigskip

{\center{\bf Zero Temperature}}
\bigskip

To calculate
$\langle in|\hat\theta^1 (t,x)\hat\theta^1 (t',x')|in\rangle$
in the `out' region, let us expand $\hat \theta_1$ in the `out' basis
\begin{equation}\label{eq:expk}
 \hat \theta^1 (t,x) = \int_{-\infty}^{+\infty} dk \left[ \hat a_k^{out} u_k^{out}(t,x)
 + h.c.\right]\ ,\end{equation}
and use the Bogoliubov transformation (\ref{eq:relationinoutopk}).
This gives
\begin{eqnarray}\label{twopoint}
 &\langle in| \{ \hat\theta^1 (t,x),\hat\theta^1 (t',x')\}|in\rangle =&  \\
 &\frac{\hbar}{2} \int_{-\infty}^{+\infty} dk
 \Big[ \left(\alpha^* u_k^{out}-\beta u_{-k}^{out *}\right)(t,x)
   \left(\alpha u_k^{out*}-\beta^*u_{-k}^{out }\right)(t',x') + c.c.\Big] & =\\
   &
 -\frac{\hbar mc_{out}}{4\pi n} \Big[ {\frac{(c_{in}^2 - c_{out}^2)}{2c_{in}c_{out}}}\ln\Delta_{uv'}^{out}\Delta_{u'v}^{out}
 +\frac{(c_{in}^2+c_{out}^2)}{2c_{in}c_{out}} \ln\Delta_{uu'}^{out}\Delta_{vv'}^{out}
 \Big]&
 \ ,
\end{eqnarray}

where we have defined
\begin{eqnarray}
&&\Delta_{uv'}^{out}\equiv (v+c_{out})t+(c_{out}-v)t'-(x-x'),\
\Delta_{u'v}^{out}\equiv \Delta_{uv'}^{out}(t,x \leftrightarrow t',x')
\ ,\ \ \ \ \ \ \ \ \ \ \ \ \ \ \ \ \nonumber \\
&&\Delta_{uu'}^{out}\equiv(v+c_{out})(t-t')-(x-x'),\  \Delta_{vv'}^{out}= (c-v_{out})(t-t')+(x-x') \ .
\end{eqnarray}

 The quantity of interest in the experiments is the normalized, one-time density-density correlation function \begin{equation} G^{(2)} (t;x-x')= \frac{1}{n^2}\langle in| \{ \hat n^1(t,x), \hat n^1 (t',x')\} |in \rangle |_{t=t'}\ ,\end{equation} that using (\ref{nm}) can be calculated by performing derivatives from the above results
 \begin{equation}
 \label{gtwo}
G^{(2)}=
\left(\frac{1}{mc_{out}^2}\right)^2\left [\partial_t\partial_{t'} +v\partial_t\partial_{x'}+v\partial_{t'}\partial_x
+v^2\partial_x\partial_{x'}\right] \langle in|  \{\hat \theta^1(t,x), \hat \theta^1 (t',x')\}|in \rangle |_{t=t'}\ .\ \ \ \ \ \ \ \
\end{equation}
 The computation is straightforward and
 we get the following contribution
\begin{equation}\label{gtwoto}
\begin{array}{cc}
G^{(2)}=
\frac{\hbar}{4\pi nmc_{out}}\left[ \frac{c_{in}^2-c_{out}^2}{2c_{in}c_{out}} \left( \frac{1}{(2c_{out}t -(x-x'))^2} + \frac{1}{(2c_{out}t -(x'-x))^2}\right)
- \frac{c_{in}^2+c_{out}^2}{c_{in}c_{out}} \frac{1}{(x-x')^2}
\right]\ .
\end{array}
\end{equation}
The interesting features are in the time dependent contributions, coming from the $\alpha\beta$ terms in (\ref{twopoint}) \footnote{The static term is instead proportional to $|\alpha|^2+|\beta|^2=1+2|\beta|^2$.}, and whose physical interpretation follows that of Eq. (\ref{vacio}). At $t=0$ and everywhere in space correlated pairs of particles with opposite momentum are created out of the vacuum state,
with velocities $v-c_{out}$ (left-moving) and $v+c_{out}$ (right-moving). At time $t$ such particles are separated by a distance $|x-x'|=2c_{out}t$, which is indeed the correlation displayed in (\ref{gtwoto}) \footnote{In the cosmological setting, the presence of the same features was noted in \cite{cp2}.}. As emphasized in \cite{ftemp}, the fact that the corresponding peak is infinite is due to the sudden jump approximation  (step-like discontinuity) employed in this analysis. A way to smooth out the transition from the `in' to the `out' regions within a time interval $\sigma_t$ and produce a finite quantity which is in good agreement with the numerical results is to multiply the integrand of Eq. (\ref{twopoint}) by $e^{-\frac{k}{k_{max}}}$, which introduces a cutoff at $k_{max}\sim \frac{1}{c\sigma_t}$. This means that the relevant physics is in the low $k$ ($<k_{max}$) modes for which the sudden transition is a good approximation to the real (smooth) situation.

\bigskip
{\center{\bf Finite temperature}}
\bigskip

We shall now consider the case in which the initial state is not the vacuum $|in\rangle$ but thermal, e.g., it is described by a thermal density matrix at temperature $T$
\begin{equation}
\rho_{th}=\prod_w (1-e^{-\hbar w/k_BT}) \sum_{N=0}^{+\infty} e^{-N\hbar w/k_BT}|N_w^{in}\rangle\langle N_w^{in}|\ .
\end{equation}
Therefore we shall need to calculate
\begin{equation}\label{trace}
Tr  \rho_{th} \hat \theta^1(t,x)\hat \theta^1(t',x')\ .
\end{equation}
To perform this calculation, starting from the decomposition (\ref{eq:expk}) and the Bogoliubov transformation
(\ref{eq:relationinoutopk}) we need now to transform to  the `in' $w$ basis
\begin{eqnarray}\label{kwin}
\hat \theta^1 = \int_0^{+\infty} dw \Big[&&\frac{\alpha^* u_{k=w/(c_{in}+v)}^{out} - \beta u_{k=-w/(c_{in}+v)}^{out*}}{{\sqrt{v+c_{in}}}}\ \hat a_w^{u, in}  \nonumber \\ +&& \frac{\alpha^* u_{k=-w/(c_{in}-v)}^{out}-\beta u_{k=w/(c_{in}-v)}^{out*} }{\sqrt{c_{in}-v}}\ \hat a_w^{v, in} + h.c.\Big]\ , \ \ \ \ \ \ \ \ \ \ \
\end{eqnarray}
where, we remind, the $k>0$ and $k<0$ `out' modes are
\begin{equation}\label{expmod}u_{k (>0)}^{out}=\sqrt{\frac{mc_{out}}{4\pi|k|n}}e^{-ik(v+c_{out})t+ikx},\ u_{k (<0)}^{out}=\sqrt{\frac{mc_{out}}{4\pi|k|n}}e^{-ik(v-c_{out})t+ikx}\end{equation}
and $\hat a_w^{u(1)}= \frac{\hat a_{k(>0)}^{(1)}}{\sqrt{c_1+v}}$, $\hat a_w^{v(1)}= \frac{\hat a_{k(<0)}^{(1)}}{\sqrt{c_1-v}}$.
Inserting (\ref{kwin}) into (\ref{trace}) as an intermediate step we get the following expression
\begin{eqnarray}
Tr \rho_{th} \hat \theta^1 \hat \theta^1 = \frac{\hbar}{2} \int_0^{+\infty}  dw
 (1+2|\beta_T|^2)
\Big( F(t,x)F^*(t',x')+ G(t,x),G^*(t',x') + c.c. \Big) ,\ \ \ \ \ \ \ \ \nonumber
\end{eqnarray}
where $|\beta_T|^2= \frac{1}{e^{\frac{\hbar w}{k_BT}}-1}$,
$F\equiv \frac{\alpha^* u_{k=w/(c_{in}+v)}^{out} - \beta u_{k=-w/(c_{in}+v)}^{out*}}{{\sqrt{v+c_{in}}}}$ and \\
$G\equiv \frac{\alpha^* u_{k=-w/(c_{in}-v)}^{out}-\beta u_{k=w/(c_{in}-v)}^{out*} }{\sqrt{c_{in}-v}}.$

Comparing with the zero-temperature case (\ref{twopoint}) we see that the effect of the initial thermal population is to stimulate the existing correlations through the factor $1+2|\beta_T|^2$, and not to create new ones.
Writing out explicitly the modes, from (\ref{expmod}),
and evaluating the integrals
we obtain the final expression
\begin{eqnarray}\label{traceintf}
&&Tr \rho_{th} \hat \theta^1 (t,x)\hat \theta^1 (t',x')=
-\frac{\hbar mc_{out}}{4\pi n} \left\{\frac{c_{in}^2-c_{out}^2}{2c_{in}c_{out}}
\ln\left[ \sinh\left(\frac{\pi k_B T}{\hbar}\frac{\Delta_{uv'}^{out}}{c_{in}-v}\right)
\sinh\left(\frac{\pi k_B T}{\hbar}\frac{\Delta_{u'v}^{out}}{v+c_{in}}\right)\right]+\right.\nonumber \\
&&\left.\frac{c_{in}^2+c_{out}^2}{2c_{in}c_{out}}
\ln\left[ \sinh\left(\frac{\pi k_BT}{\hbar}\frac{\Delta_{uu'}^{out}}{v+c_{in}}\right)
\sinh\left(\frac{\pi k_BT}{\hbar}\frac{\Delta_{vv'}^{out}}{c_{in}-v}\right)\right]
 \right\} \ .
\end{eqnarray}
From this result the calculation of $G^{(2)}_{T}$, using (\ref{gtwo}), gives
\begin{eqnarray}\label{gtwot}
&&G^{(2)}_{T}(t;x-x')=\frac{\hbar}{4\pi nmc_{out}}\left\{-\frac{c_{in}^2+c_{out}^2}{2c_{in}c_{out}}\left[ \frac{A^2}{\sinh^2[A(x-x')]}+\frac{B^2}{\sinh^2[B(x-x')]} \right] +\right.\nonumber\\
&&\left.\frac{c_{in}^2-c_{out}^2}{4c_{in}c_{out}} \left[A^2\left(\frac{1}{\sinh^2[A(2c_{out} t-(x-x'))]}+ \frac{1}{\sinh^2[A(2c_{out} t+(x-x'))]}\right) +\right.\right.\nonumber \\
&&\left.\left.B^2\left(\frac{1}{\sinh^2[B(2c_{out} t-(x-x'))]}+ \frac{1}{\sinh^2[B(2c_{out} t+(x-x'))]}\right)\right]  \right\}\ ,
\end{eqnarray}
where $A\equiv \frac{\pi k_B T}{\hbar(v+c_in)}$ and $B\equiv \frac{\pi k_B T}{\hbar(c_in-v)}$.
Comparing with (\ref{gtwoto}), we see that indeed the peaks structure is the same, and in particular the time-dependent ones receive the following finite contribution
\begin{equation}
\Delta G^{(2)}_{T, |x-x'|=2c_{out}t}= -\frac{\pi (k_BT)^2}{12nmc_{out}}\frac{(c_{in}^2-c_{out}^2)}{2c_{in}c_{out}} \frac{(v^2+c_{in}^2)}{(v+c_{in})^2(c_{in}-v)^2}\ ,
\end{equation}
which in the more realistic case of a transition within a time $\sigma_t$ between the `in' and the `out' regions
and $v=0$ has been shown to be a good approximation to the numerical results for low temperatures $T<\hbar/k_B\sigma_t$ \cite{ftemp}.\\
Finally, it is noted in section IV.A of \cite{nostrodisp} that dispersion effects do not change qualitatively the hydrodynamical features discussed here.

\bigskip
{\subsection{\bf Simulation of the formation of non-homogeneous backgrounds using step-like discontinuities}}
\bigskip

\begin{figure}[htbp]
\begin{center}

\resizebox{!}{6cm}{\includegraphics{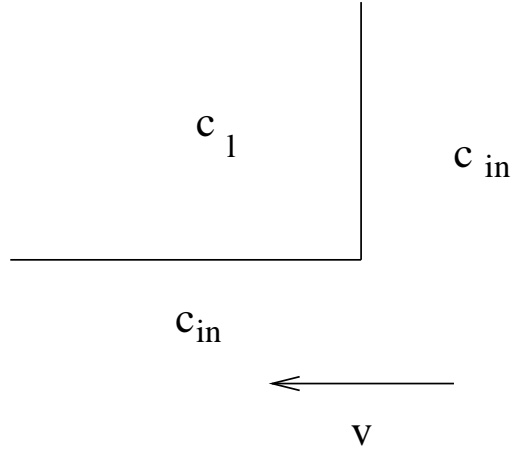}}

\caption{Temporal formation of a spatial step-like discontinuity ($c_r=c_{in}$).}
\label{figure:tempspatgrad}
\end{center}
\end{figure}

We shall now combine the analysis of the two previous subsections to construct more complicated time-dependent backgrounds. Consider an initial infinite homogeneous condensate with sound speed $c_{in}$ turning into a nonhomogeneous configuration well described by two semiinfinite (left and right) homogeneous condensates with sound speeds, respectively, $c_l$  and $c_r$. Technically the model we use consists of a discontinuity in $t$ followed by a discontinuity in $x$, i.e.
$c(t,x)=c_{in}\theta(-t) + \theta(t) (c_l\theta(-x) + c_r\theta(x))$. For simplicity we shall consider
the case in which $c_{in}=c_r$, represented in Fig.\ref{figure:tempspatgrad}.

Such a construction is the closest, in spirit, to the time-dependent formation of an acoustic black hole as considered in \cite{fteo}, \cite{fnum}, the important difference being, of course, that our hydrodynamic analysis can only be made for subsonic flows. In addition, a realistic realization of this scenario requires non vanishing transition regions in both $t$ and $x$ of width $\sigma_t$ and $\sigma_x$.

To solve (\ref{sde})  for all $x$ and $t$, first we write down the general solutions in the three different regions , `in' ($t<0$) and `out' left ($t>0,\ x<0$) and right ($t>0,\ x>0$), and then impose the matching conditions (\ref{mcondsecb}) and (\ref{mcondsec}).
The former are performed at fixed $k$, while the latter at fixed $w$. Therefore both $k$ and $\omega$ decompositions (\ref{decompk}) and (\ref{decompw}) will be needed, the respective modes and operators being related by (see also \cite{cp1})
\begin{eqnarray}\label{relkw}
u_{k(<0)}=\sqrt{c-v}\ u_w^v\ &,& \ u_{k(>0)}=\sqrt{c+v}\ u_w^u\\
\hat a_{k(<0)}=\sqrt{c-v}\ \hat a_w^{v}\ &,& \ \hat a_{k(>0)}=\sqrt{c+v}\ \hat a_w^{u}\ .\end{eqnarray}

To solve (\ref{mcondsecb}) at $t=0$ consider the general solutions of (\ref{sde}) at fixed $k$ in the `in' region
\begin{equation}
\theta^1_{in}= e^{ikx}\sqrt{\frac{mc_{in}}{4\pi |k| n}} \left[ A_v^{in}e^{-ik(v-c_{in})t}  + A_u^{in}e^{-ik(v+c_{in})t}  \right],
\end{equation}
and in the `out' left and right regions
\begin{eqnarray}
\theta^1_{l}(k)= e^{ikx}\sqrt{\frac{mc_{l}}{4\pi |k| n}} \left[ A_v^{l}e^{-ik(v-c_{l})t}  + A_u^{l}e^{-ik(v+c_{l})t}  \right]\ , \\
\theta^1_{r}(k)= e^{ikx}\sqrt{\frac{mc_{in}}{4\pi |k| n}} \left[ A_v^{r}e^{-ik(v-c_{in})t}  + A_u^{r}e^{-ik(v+c_{in})t}  \right]\ .
\end{eqnarray}
For $x<0$ the `in' and $l$ amplitudes are related through the Bogoliubov matrix
\begin{equation}
\label{xneg}
     \left( \begin{array}{c}
       A_v^{l} \\
       A_u^{l}  \\
     \end{array} \right)
   = \frac{1}{2\sqrt{c_{in} c_l}}\left(
            \begin{array}{cc}
               c_{in}+c_l &
               c_{in}-c_l \\
               c_{in}-c_l &
               c_{in}+c_l \\ \end{array} \right) \left(
                \begin{array}{c}
                  A_v^{in}  \\
                  A_u^{in} \\
                \end{array}
              \right) \ .
\end{equation}

At $x>0$ the matching is trivial because we chose $c_{in}=c_r$, therefore
\begin{equation}
A_v^{r}=A_v^{in},\ A_u^{r}=A_u^{in}\ .\end{equation}

The matching equations (\ref{mcondsec}) at $x=0$ for all $t>0$ relate the general solutions at fixed $w$ in the $l$ and $r$ regions
\begin{eqnarray}
\theta^{1}_l(\omega) &=& e^{-iwt}\sqrt{\frac{mc_l}{4\pi wn}}
 \left[ e^{i\frac{\omega }{v-c_l}x} A_v^{l}+
e^{i\frac{\omega }{v+c_l}x}  A_u^{l} \right] \ ,  \\
 \theta^{1}_r(\omega) &=&
e^{-iwt}\sqrt{\frac{mc_{in}}{4\pi wn}}
\left[ e^{i\frac{\omega }{v-c_{in}}x}  A_v^{r}+
e^{i\frac{\omega }{v+c_{in}}x}  A_u^{r} \right]  \end{eqnarray}
via the scattering matrix
\begin{equation}
\label{scattmat}
     \left( \begin{array}{c}
       A_v^{l} \\
       A_u^{l}  \\
     \end{array} \right)
   = \frac{1}{2\sqrt{c_r c_l}}\left(
            \begin{array}{cc}
               c_{in}+c_l &
               c_{in}-c_l \\
               c_{in}-c_l &
               c_{in}+c_l \\ \end{array} \right) \left(
                \begin{array}{c}
                  A_v^{r}  \\
                  A_u^{r} \\
                \end{array}
              \right) \ .
\end{equation}


In the next subsection we shall construct the explicitly the `in' modes and study their time evolution in the $t>0$ region. This analysis is also useful to clarify what happens at the point $x=t=0$, belonging to both the spatial and the temporal step-like discontinuities, and to the boundary conditions satisfied as one approaches it.

\bigskip
{\subsubsection{\bf  Relating the `in' and  `out' basis}}
\bigskip

\begin{figure}[htbp]
\begin{center}

\resizebox{!}{6cm}{\includegraphics{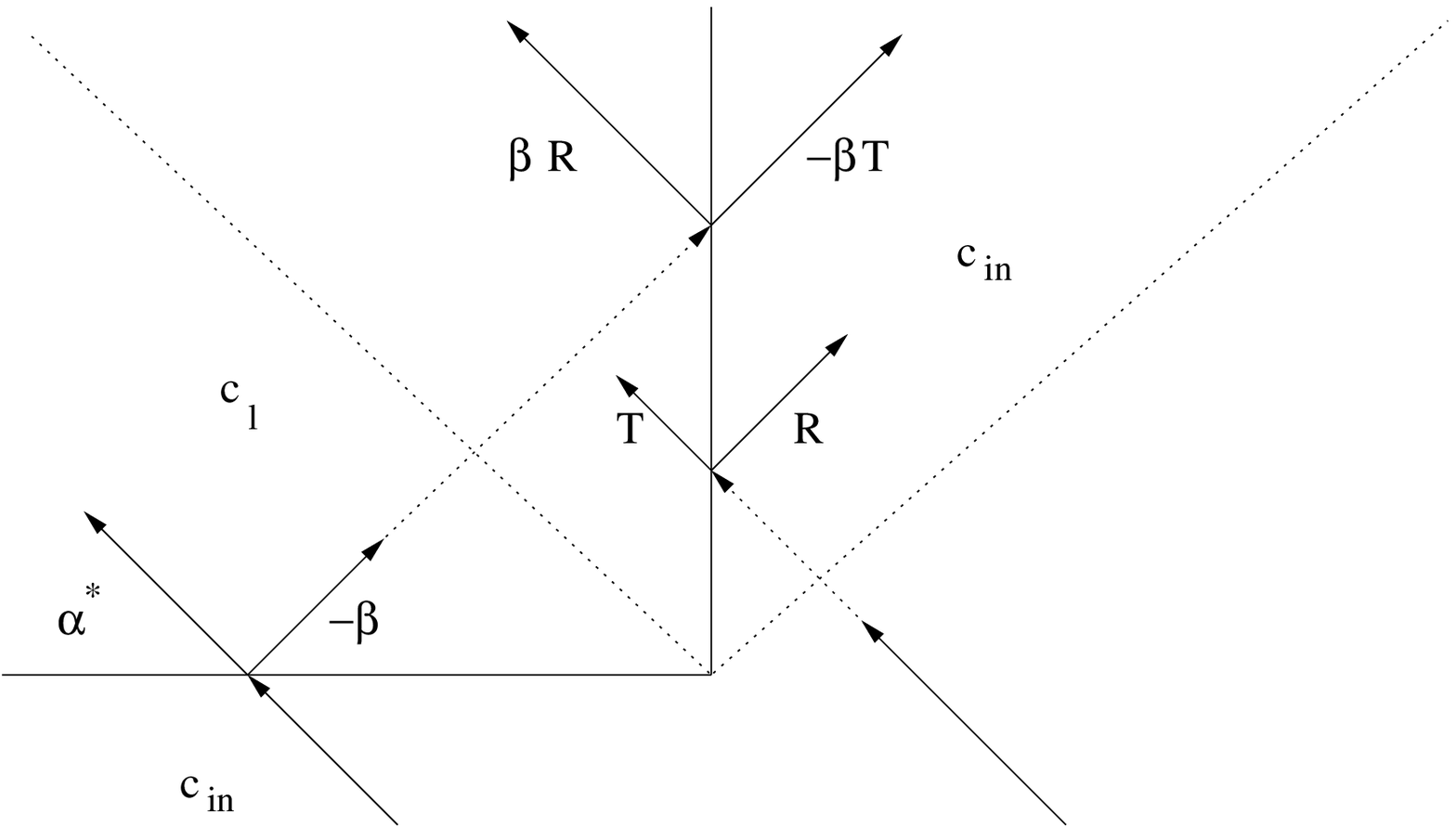}}

\caption{Evolution of $u_k^{in}(k<0)$ modes.}
\label{figure:vhidro1}
\end{center}
\end{figure}

\begin{figure}[htbp]
\begin{center}

\resizebox{!}{6cm}{\includegraphics{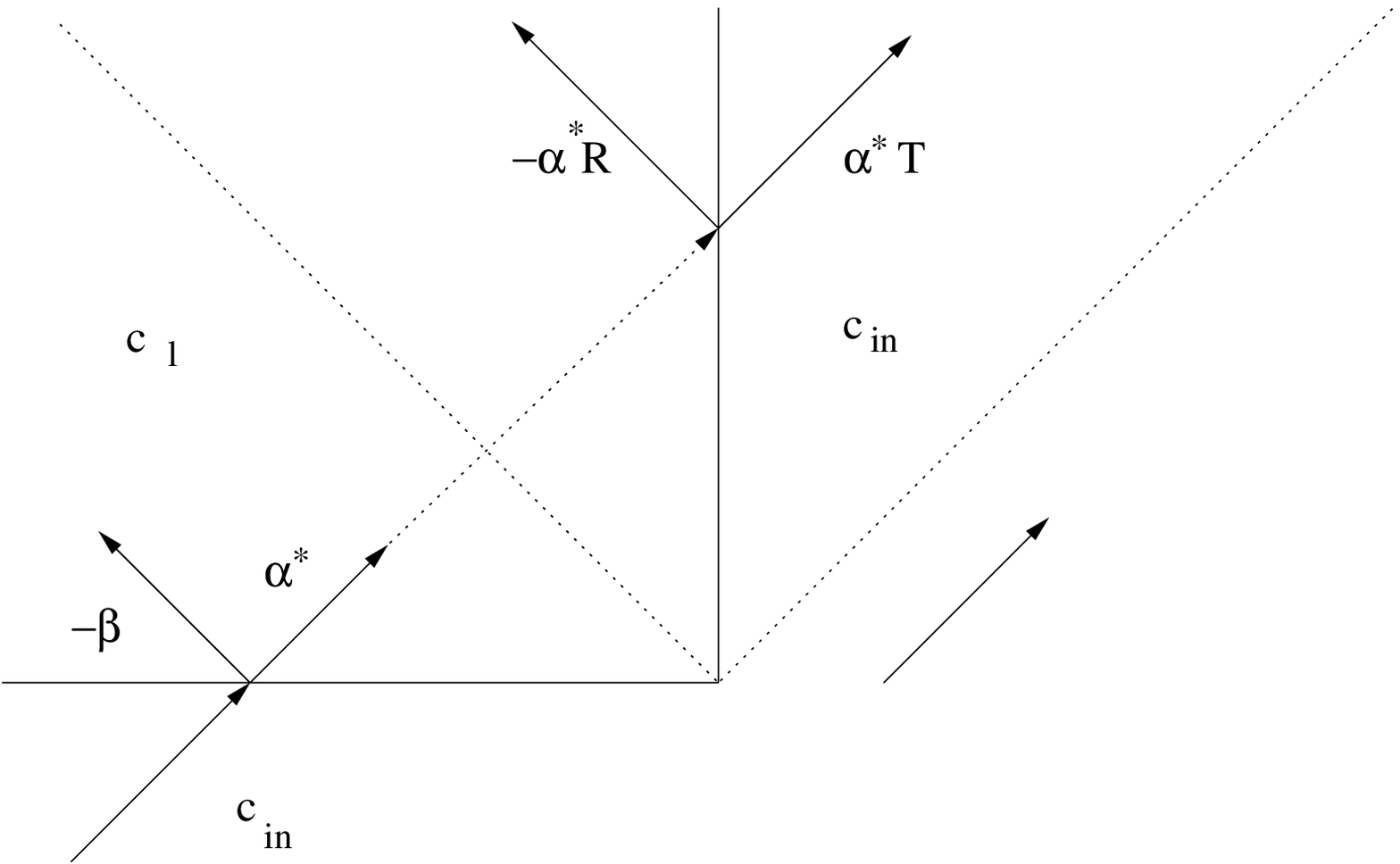}}

\caption{Evolution of $u_k^{in}(k>0)$ modes.}
\label{figure:uhidro1}
\end{center}
\end{figure}

The way the `in' $k$ basis
is related to the `in' and 'out' $w$ basis ($t>0$)
is shown in Figs. \ref{figure:vhidro1} and \ref{figure:uhidro1},
where the Bogoliubov and scattering coefficients are
\begin{equation}
\alpha^*=\frac{c_{in}+c_l}{2\sqrt{c_{in}c_l}},\ -\beta=\frac{c_{in}-c_l}{2\sqrt{c_{in}c_l}},\
R=-\frac{c_{in}-c_l}{c_{in}+c_l},\ T=\frac{2\sqrt{c_{in}c_l}}{c_{in}+c_l}.\end{equation}
These pictures are constructed by combining those of Figs. \ref{figure:modosin} and \ref{figure:tempgrad}
for the two types of discontinuities.

\bigskip
{\center{\bf Left-moving `in' modes $u_{k(<0)}^{in}$}}
\bigskip

The left moving `in' modes produce at $t= 0^+$ three different components, two due to the discontinuity on the $x<0$ side and
the third, freely propagating at $x>0$.
Only two of them scatter at the $x=0$ discontinuity, the other one being disconnected from it (it propagates outside the future sound cone of the origin).
Explicitly,
\begin{eqnarray}\label{evolvv}
&&\sqrt{c_{in}}\ e^{-ik(v-c_{in})t+ikx} \to \sqrt{c_l} \{   \theta(\frac{x}{v-c_l}-t)\alpha^* e^{-ik(v-c_l)t+ikx} -\beta e^{-ik(v+c_l)t+ikx}
  \nonumber \\\
&&+\theta(t-\frac{x}{v-c_l})[\beta R^* e^{-ik(v+c_l)t+ik\frac{v+c_l}{v-c_l}x} +Te^{-ik(v-c_{in})t+ik\frac{v-c_{in}}{v-c_l}x}]\}
+ \sqrt{c_{in}}\{ e^{-ik(v-c_{in})t+ikx} \nonumber \\\
&&+\theta(t-\frac{x}{v+c_{in}}) [ -\beta T^* e^{-ik(v+c_l)t+ik\frac{v+c_l}{v+c_{in}}x} +Re^{-ik(v-c_{in})t+ik\frac{v-c_{in}}{v+c_{in}}x}]\}  \ .
\end{eqnarray}
We see that in the $t>0$ region there are several ways to get to the point $x=t=0$. If we approach it from outside its sound cone from the left (the right part being trivial) we have the usual components arising from the temporal step-like discontinuity satisfying the boundary conditions (\ref{mcondsecb}).
If instead we approach it from inside the sound cone the existing components there satisfy the boundary conditions  (\ref{mcondsec}).
Note that the mode (\ref{evolvv}) is continuous at $x=t=0$, irrespective of how we get to it, and along the sound cone generators  $t=\frac{x}{v-c_l},\ t=\frac{x}{v+c_{in}}$ as one can easily verify from  the relations $\alpha^*=\beta R^* + T$ and  $R-\beta T^*=0$.

In the next subsection we shall perform the analysis of the correlations patterns and look for stationary signals (like the Hawking signal, see \cite{fteo}).  For this reason we shall focus our attention  to the interior of the future sound cone of the origin where we have
\begin{eqnarray}\label{evolv}
&&u_{k(<0)} \to  -\beta \sqrt{v+c_l}\ u^{in, u*}_{w=-k(v+c_l)} +\sqrt{c_r -v}\ u^{v,in}_{w=(v-c_r)k}=\nonumber \\
&&-\beta \sqrt{v+c_l}( -R^*u^{v,out *}_{w=-k(v+c_l)}+T^*u^{u,out *}_{w=-k(v+c_l)})
+\sqrt{c_r -v}( Tu^{v,out}_{w=(v-c_r)k}+Ru^{u,out}_{w=(v-c_r)k})\ .
\end{eqnarray}
We mention that with this restriction we are neglecting transient effects such as those arising from the temporal step-like discontinuity (see (\ref{gtwoto})), see the numerical results in \cite{fnum}.
Such effects are absent also in \cite{fteo}  because there we considered the conformal field approximation for $\theta^1$.

\bigskip
{\center{\bf Right-moving `in' modes $u^{in}_{k(>0)}$}}
\bigskip

In this case of the three components produced at $t=0^+$  only the one proportional to $\alpha^*$ propagates
inside the future light-cone of the origin and scatters at the spatial step-like discontinuity. The explicit time evolution is
\begin{eqnarray}\label{evolvv}
&&\sqrt{c_{in}} e^{-ik(v+c_{in})t+ikx} \to \sqrt{c_l} \{   \theta(\frac{x}{v-c_l}-t)(-\beta) e^{-ik(v-c_l)t+ikx}+\alpha^* e^{-ik(v+c_l)t+ikx}
  \nonumber \\\
&&
+\theta(t-\frac{x}{v-c_l})(-\alpha^* R) e^{-ik(v+c_l)t+ik\frac{v+c_l}{v-c_l}x} \}+ \sqrt{c_{in}}\{  \theta(t-\frac{x}{v+c_{in}}) \alpha^*T
e^{-ik(v+c_l)t+ik\frac{v+c_l}{v+c_{in}}x} \nonumber \\\
&&+\theta(\frac{x}{v+c_{in}}-t)  e^{-ik(v+c_{in})t+ikx} \}  \ .
\end{eqnarray}
The relations $-\beta=-\alpha^*R, \ \alpha^*T=1$ guarantee continuity at $x=t=0$ and along the sound cone generators $t=\frac{x}{v-c_l},\ t=\frac{x}{v+c_{in}}$. The boundary conditions involving derivatives  in (\ref{mcondsecb}) and  (\ref{mcondsec}) are satisfied at the origin by approaching it from, respectively, outside (from the left) and inside its sound cone.

Restricting again to the interior of the sound cone of $x=t=0$ we have
\begin{equation}\label{evolu}
u^{in}_{k(>0)}\to \alpha^*\sqrt{v+c_l}\ u^{u,in}_{w=(v+c_l)k}=
\alpha^*\sqrt{v+c_l}(-Ru_{w=(v+c_l)k}^{v,out}+Tu^{u,out}_{w=(v+c_l)k})\ .\end{equation}

\bigskip
\bigskip
Let us now construct the explicit connection between the $\hat a,\ \hat a^{\dagger}$ operators of `in' $k$ decomposition ($t<0$)
 \begin{equation}\label{uuu}
\hat \theta^1 (t,x) =\int_{0}^{+\infty} dk \left[ \hat a_k^{in} u_k^{in} (t,x)  + \hat a_{-k}^{in \dagger} u_{-k}^{in}(t,x) + h.c. \right]\ , \end{equation}
where we have explicitly separated the positive $k$ (left-moving) and the negative $k$ (right-moving) contributions, and the `in' and `out' $w$ ones in the $t>0$ region
\begin{equation}
\hat \theta^1 (t,x) =\int_0^{+\infty} dw \left[ \hat a_w^{v,in(out)} u^{v,in(out)}_w (t,x)+ \hat a_w^{u,in(out)} u^{ u,in(out)}_w (t,x)  + h.c. \right]\ .
\end{equation}
By inserting the first lines of eqs. (\ref{evolv}), (\ref{evolu}) into (\ref{uuu}) and changing from $k$ to $w$ we have
\begin{equation}\label{lucu}
\hat \theta_1 = \int_0^{+\infty} dw \Big[\frac{ (\alpha^* \hat a_{k} - \beta^* \hat a_{-k}^{\dagger})|_{k=\frac{w}{c_{l}+v}}}{{\sqrt{v+c_{l}}}}\  u_w^{u, in}  +\frac{\hat a_{-k=\frac{w}{v-c_{in}}}}{\sqrt{c_{in}-v}}\ u_w^{v,in} + h.c.\Big]\ , \ \ \ \ \ \ \ \ \ \ \
\end{equation}
from which
\begin{equation}\label{yyy}
\hat a_w^{u,in}=\frac{ (\alpha^* \hat a_{k} - \beta^* \hat a_{-k}^{\dagger})|_{k=\frac{w}{c_{l}+v}}}{\sqrt{c_l+v}}\ ,
\ \ \hat a_w^{v,in}=\frac{a_{-k=\frac{w}{v-c_{in}}}}{\sqrt{c_{in}-v}}\ .
\end{equation}
To change from the `in' to the 'out' $w$ basis we can use the second lines of (\ref{evolv}), (\ref{evolu}) into (\ref{uuu}) or, directly, (\ref{eq:relationinoutop}) and (\ref{yyy}) :
\begin{eqnarray}\label{outwink}
\hat a_w^{u,out} &=& R \frac{a_{-k=\frac{w}{v-c_{in}}}}{\sqrt{c_{in}-v}} +T \frac{ (\alpha^* \hat a_{k} - \beta^* \hat a_{-k}^{\dagger})|_{k=\frac{w}{c_{l}+v}}}{\sqrt{c_l+v}}\ ,\nonumber\\
\hat a_w^{v,out} &=& T\frac{a_{-k=\frac{w}{v-c_{in}}}}{\sqrt{c_{in}-v}} -R\frac{ (\alpha^* \hat a_{k} - \beta^* \hat a_{-k}^{\dagger})|_{k=\frac{w}{c_{l}+v}}}{\sqrt{c_l+v}}\ .\end{eqnarray}

\bigskip
{\subsubsection{\bf Correlations }}
\bigskip

To evaluate in the $t>0$ region the two-point correlation function of $\hat \theta_1$ in the `in' vacuum
we can use the `out' $w$ decomposition
\begin{equation}
\hat \theta^1 (t,x) =\int_0^{+\infty} dw \left[ \hat a_w^{v,out} u^{v,out}_w (t,x)+ \hat a_w^{u,out} u^{ u,out}_w (t,x)  + h.c. \right]\
\end{equation}
and then the relations (\ref{outwink}). This gives
\begin{eqnarray}\label{prim}
&& \langle in|\hat\theta^1 (t,x)\hat\theta^1 (t',x')|in\rangle = \frac{\hbar}{2} \int_0^{+\infty}
\Big[ u_w^{v,out}(x,t)u_w^{v,out *}(x',t')+ u_w^{u,out}(x,t)u_w^{u,out *}(x',t')+\nonumber\\
&&2|\beta|^2 (-T^*u_w^{u,out*} +R^*u_w^{v,out*})(t,x)(-Tu_w^{u,out}+Ru_w^{v,out})(t',x') +c.c.\Big]\ .
\end{eqnarray}
Alternatively, we can use the `in' $w$ decomposition
\begin{equation}
\hat \theta^1 (t,x) =\int_0^{+\infty} dw \left[ \hat a_w^{v,in} u^{v,in}_w (t,x)+ \hat a_w^{u,in} u^{ u,in}_w (t,x)  + h.c. \right]\ ,
\end{equation}
from which using (\ref{yyy})
\begin{equation}\label{secon}
\langle in|\hat\theta^1 (t,x)\hat\theta^1 (t',x')|in\rangle = \frac{\hbar}{2} \int_0^{+\infty} dw\Big[
u_w^{v,in}(x,t)u_w^{v,in *}(x',t')+ (1+2|\beta|^2)u_w^{u,in}(x,t)u_w^{u,in *}(x',t') +c.c. \Big]\ .
\end{equation}
Given the relations (\ref{eq:relationinoutmo}) between `in' and `out' $w$
modes, the two expressions (\ref{prim}) and (\ref{secon}) are clearly equivalent. With respect to the stationary case (\ref{tupi}) ($\beta=0$), we see in (\ref{secon}) the different weight the `in' $v$ and $u$ modes have, modifying the correlation picture. In particular, considering points respectively in the left (say, $x<0$) and the right ($x'>0$) region we have (see (\ref{uvin}))
$$\langle in|\hat\theta^1 (t,x)\hat\theta^1 (t',x')|in\rangle=
\frac{\hbar}{2} \int_0^{\infty}d\omega \Big[ T^*(1+2|\beta|^2)u_w^{u,l}(t,x)u_w^{u,r *}(t',x')+$$
$$Tu_w^{v,l}(t,x)u_w^{v,r *}(t',x') +  \Big(R^*T-RT^*(1+2|\beta|^2)\Big)u_{\omega}^{v,l}(t,x)u_{\omega}^{u,r *}(t',x') + c.c. \Big] =$$
$$-\frac{\hbar m}{4\pi n}\Big[ \frac{(c_{in}^2+c_l^2)}{c_{in}+c_l}
\ln (t-t'-\frac{x}{v+c_l}+\frac{x'}{v+c_{in}})
+\frac{2c_{in}c_l}{c_{in}+c_l}
\ln(t-t' -\frac{x}{v-c_l}+\frac{x'}{v-c_{in}}) $$
$$+ \frac{(c_{in}-c_l)^3}{(c_{in}+c_l)^2}\ln (t-t'-\frac{x}{v-c_l}+\frac{x'}{v+c_{in}})
\Big] \ ,$$
where, comparing with (\ref{eq:twopoint1}), we see here the different contributions coming from the $uu$ and $vv$
sectors and the appearance of a new term, a nonvanishing $uv$ correlator proportional to $(RT^*)|\beta|^2$.

We can also compute the left-right density-density function
\begin{equation} G^{(2)}_{lr}
=
\frac{1}{(mc_{in}c_l)^2}
\left[\partial_t\partial_{t'} +v\partial_t\partial_{x'}+v\partial_{t'}\partial_x
+v^2\partial_x\partial_{x'}\right] \langle in|  \{\hat \theta_1(t,x), \hat \theta_1 (t',x')\}|in \rangle |_{t=t'}
\end{equation}
which gives the following contribution
$$G^{(2)}_{lr}=
-\frac{\hbar }{4\pi nmc_{in}c_l(c_{in}+c_l)}\Big[ \frac{(c_{in}^2+c_l^2)}{(v+c_l)(v+c_{in})}
\frac{1}{(\frac{x}{v+c_l}-\frac{x'}{v+c_{in}})^2}$$
$$+\frac{2c_{in}c_l}{(v-c_l)(v-c_{in})}\frac{1}{
(\frac{x}{v-c_l}-\frac{x'}{v-c_{in}})^2}
- \frac{(c_{in}-c_l)^3}{(c_{in}+c_l)(v-c_l)(v+c_{in})}\frac{1}{(\frac{x}{v-c_l}-\frac{x'}{v+c_{in}})^2}
\Big] \ .$$
The last term, the $uv$ correlator, has a peak (divergence) for $x<0,\ x'>0$.
This feature is present in the numerical simulations of time-dependent subsonic configurations performed (together with the formation of acoustic black holes) in \cite{fnum}.
As already remarked,
to have finite results we would need more realistic configurations with smooth transitions of width $\sigma_t$, $\sigma_x$ instead of the sudden jumps considered in our analysis. A way to take into account this is to introduce a cutoff in our calculation. Actually in this model, due to the two types of discontinuities we have two such cutoffs, $\frac{1}{\sigma_t}$ and  $\frac{c}{\sigma_x}$.

\bigskip
{\section{\bf Hawking radiation and discussion}}
\bigskip

It is interesting, at this point, to ask whether step-like discontinuities are suitable to study the Hawking effect in correlations and to which extent.

In gravitational physics, the Hawking effect \cite{h} arises from vacuum fluctuations in the dynamical background
of a collapsing star forming a black hole. The universality of the emitted thermal radiation (with temperature $T_H=\frac{\hbar\kappa}{2\pi k_B}$, where $\kappa$ is the surface gravity at the horizon) allows us to use simple analytical models (for instance shock waves) to mimic the time dependence in a real process (see for instance \cite{sandropepe}).
The mechanism by which the black hole emits particles is understood in terms of pair creation in the outgoing ($u$) sector in the near-horizon region, one member outside the horizon (Hawking quanta) and the other inside (the partner, trapped inside the black hole) \cite{cw}, \cite{mp}. This produces a characteristic signal, a finite stationary peak in  correlators (for instance, in the point-split stress tensor) proportional to $\kappa^2$ for points situated on either side with respect to the horizon and in correspondence  with the positions of the created quanta.

\begin{figure}[htbp]
\begin{center}

\resizebox{!}{6cm}{\includegraphics{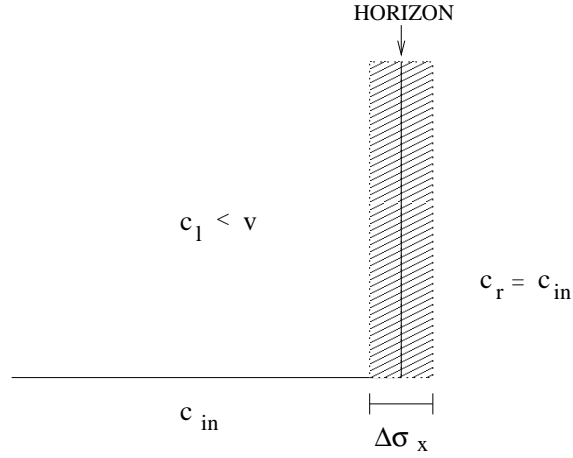}}

\caption{Formation of an acoustic black hole.}
\label{figure:paperidro}
\end{center}
\end{figure}

In our context we would need a simple analytical model allowing us to study the dynamical formation of a fluid with a region of supersonic flow out of an everywhere subsonic configuration. As already commented in section 4.1.1, step-like discontinuities in $x$ can only be defined, in the hydrodynamic limit, for subsonic configurations. The formation of an acoustic black hole with a regular surface gravity is achieved by introducing a small transition region of width $\Delta \sigma_x$ around $x=0$ where the sound velocity varies smoothly from $c_r=c_{in}$ (right region) to $c_l<|v|$ (left region), see Figure \ref{figure:paperidro}. This would give a surface gravity $\kappa \sim \frac{c}{\Delta \sigma_x}$.

The physics of the Hawking effect tells us to look for the $uu$ correlations between the pairs created on the horizon at its formation, its members following, respectively, the trajectories $x'=(c_r+v)t'\ >0$ (Hawking quanta) and $x=(c_l+v)t\ <0$  (partner). At time $t=t'$ this corresponds to points $x,x'$ such that $\frac{x}{v+c_l}-\frac{x'}{v+c_r}=0$.

\begin{figure}[htbp]
\begin{center}

\resizebox{!}{6cm}{\includegraphics{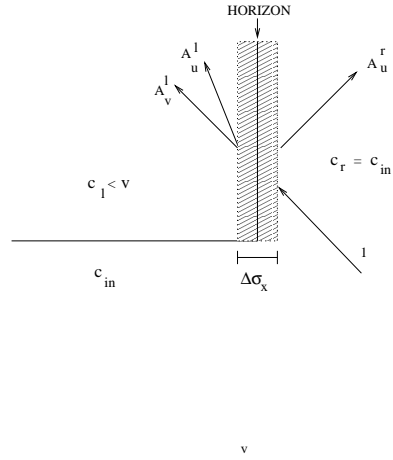}}

\caption{`in' left moving modes in an acoustic black hole.}
\label{figure:paperidro2}
\end{center}
\end{figure}

A correlation of this type is present in the evolution of the `in' left moving modes from the right, see Figure \ref{figure:paperidro2} \footnote{The main contribution to the Hawking signal comes from the evolution of `in' right-moving modes, as described in \cite{fteo}. We cannot capture it by modeling the black hole with a spatial step-like discontinuity (to get it we need a smooth transition between the subsonic and supersonic regions, see also the analysis in \cite{fab}). The type of correlation discussed here is a back-scattering type correction to it. Also, we mention that
there is another type of left-right correlation in the $uv$ channel, that exists also in the subsonic case.}. The amplitudes $A_u^r, A_u^l, A_v^l$ depend on the details of the modes propagation in the transition region. We mention that an expression for them can be found by considering step-like discontinuities in the exact theory (with $\xi \neq 0$) and taking the dispersionless limit $\xi \rightarrow 0$ at the end of the calculations: $A_u^r = \frac{c_r+v}{v-c_r}, A_u^l = \sqrt{\frac{c_r}{c_l}}\frac{v+c_l}{c_r-v}, A_v^l = \sqrt{\frac{c_r}{c_l}}\frac{c_l-v}{c_r-v}$.

We can then perform an analysis similar to that of section IV.C to consider the dynamical formation of an acoustic black hole-like configuration and taking it into account the presence of a finite $\Delta \sigma_x$ by imposing a cut-off $\omega_{max}\sim \kappa$ in the integration over frequencies. It is not difficult to show that the $G^{(2)}$ correlator has indeed a left-right $uu$ Hawking-like peak located at $\frac{x}{v+c_l}=\frac{x'}{v+c_r}$ and proportional to
\begin{equation}
G^{(2)uu}_{lr}\sim \frac{\hbar}{4\pi mn \sqrt{c_rc_l}}\frac{A_u^lA_u^{r*}}{(v+c_l)(v+c_r)}\kappa^2,
\end{equation}
that has a structure similar to that found in \cite{fteo}. The next step is to extend the analysis of this paper by considering dispersion effects ($\xi\neq 0$) as well. This is dealt with in detail in \cite{nostrodisp}.

\bigskip
{\bf Acknowledgements}: We thank R. Balbinot, I. Carusotto, R. Parentani, A. Recati and M. Rinaldi for useful discussions.
\noindent

\end{document}